\documentclass[aps,english,twocolumn]{revtex4}

\usepackage{graphicx}
\usepackage{amsmath, amssymb}
\usepackage{babel}

\begin{document}

\title{Topological characterization of quantum phase transitions
in a S=1/2 spin model}

\author{Xiao-Yong Feng$^1$, Guang-Ming Zhang$^2$, and Tao Xiang$^1$}

\affiliation{$^1$Institute of Theoretical Physics, Chinese Academy
of Sciences, P.O. Box 2735, Beijing 100080, China \\
$^2$Department of Physics, Tsinghua University, Beijing 100084, China}

\date{\today }

\begin{abstract}
We have introduced a novel Majorana representation of S=1/2 spins
using the Jordan-Wigner transformation and have shown that a
generalized spin model of Kitaev defined on a brick-wall lattice is
equivalent to a model of non-interacting Majorana fermions with
$Z_2$ gauge fields {\it without} redundant degrees of freedom. The
quantum phase transitions of the system at zero temperature are
found to be of topological type and can be characterized by
\textit{nonlocal} string order parameters. In appropriate dual
representations, these string order parameters become \textit{local}
order parameters and the basic concept of Landau theory of
continuous phase transition can be applied.
\end{abstract}

\maketitle

The Landau theory of second order phase transitions has fertilized
modern statistical and condensed matter physics. Essential is to use
local order parameters to describe the continuous phase transition
between a disordered and an ordered phase associated with symmetry
breaking.\cite{sachdev-book} However, a quantum phase transition
driven entirely by quantum fluctuations at zero temperature can
occur between two disordered phases without any symmetry
breaking.\cite{read-green,volovic-book} A typical example is the
topological phase transition between two neighboring quantum Hall
plateaus in the fractional quantum Hall effect.\cite {wen-book} As
no conventional Landau-type order parameters can be used, a
comprehensive characterization of this kind of quantum phase
transitions has become one of the most challenging issues in
condensed matter theory.

\begin{figure}[h]
\centering \includegraphics [width=7cm]{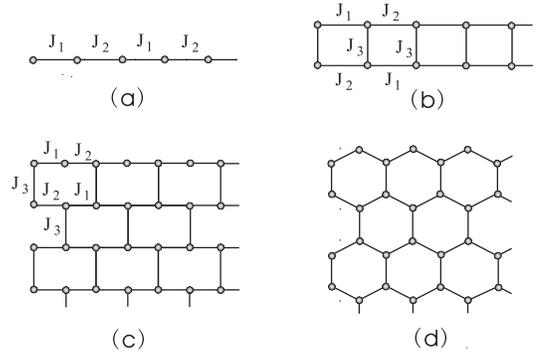} \caption{ A
brick-wall lattice (c) with its equivalent honeycomb lattice (d).
(a) and (b) are the one- and two-row limits of the brick-wall
lattices, respectively. } \label{fig:brickwall}
\end{figure}

In this work, we present a theoretical analysis of quantum phase
transitions in the following $S=1/2$ spin model first introduced
by Kitaev\cite{Kitaev}
\begin{equation}
H=\sum_{j+l=\text{even}}(J_{1}\sigma _{j,l}^{x}\sigma
_{j+1,l}^{x}+J_{2}\sigma _{j-1,l}^{y}\sigma _{j,l}^{y}+J_{3}\sigma
_{j,l}^{z}\sigma _{j,l+1}^{z}),  \label{eq:model}
\end{equation}
where $j$ and $l$ denote the column and row indices of the lattice.
This model is defined on a brick wall lattice (Fig.
\ref{fig:brickwall}c), which is an alternative representation of a
honeycomb lattice (Fig. \ref{fig:brickwall}d). Here we emphasize the
use of the brick-wall lattice because, as will be shown later, it
presents an intuitive configuration for labeling sites from which
both the Jordan-Wigner and duality transformations of spins can be
more naturally introduced. The one- and two-row limits of the
brick-wall lattices with periodic boundary conditions are simply a
single spin chain (Fig. \ref {fig:brickwall}a) and a two-leg ladder
(Fig. \ref{fig:brickwall}b), respectively. In these limits, the
analysis of the model is greatly simplified, while the results
obtained are highly nontrivial and can be extended with proper
modification to other brick wall lattices. In the work of
Kitaev\cite{Kitaev}, the two-dimensional phase diagram of the ground
state was studied. Here we will concentrate more on the analysis of
quantum phase transitions in an arbitrary brick-wall lattice. Our
studies reveal many generic features of topological quantum phase
transitions in this system. It also sheds light on the understanding
of more general systems, where analytic solutions are not available.

To solve this model, let us first introduce the following Jordan-Wigner
transformation to represent spin operators along each row with spinless
fermion operators:
\begin{equation}
\sigma_{jl}^{+}=2a_{jl}^{\dagger }e^{i\pi \left(
\sum_{i,k<l}a_{ik}^{\dagger }a_{ik}+\sum_{i<j}a_{il}^{\dagger
}a_{il}\right) }.
\end{equation}
For each pair of fermion operators $(a,a^{\dagger })$, we can
further define two Majorana fermion operators $(c,d)$:
$c_{jl}=i(a_{jl}^{\dagger} - a_{jl})$ and $d_{jl}=a_{jl}^{\dagger }
+ a_{jl}$ when $j+l = \text{even}$, and $ c_{jl}=a_{jl}^{\dagger } +
a_{jl}$ and $d_{jl}=i(a_{jl}^{\dagger }-a_{jl})$ when
$j+l=\text{odd}$. With these definitions, it is straightforward to
show that at each given row only $c$-type of Majorana fermions are
present and the Hamiltonian can be expressed as
\begin{eqnarray}
H &=& -i\sum_{j+l=\text{even}}
(J_{1}c_{j,l}c_{j+1,l}-J_{2}c_{j-1,l}c_{j,l}
\notag \\
&&\qquad\qquad + J_{3}D_{jl}c_{j,l}c_{j,l+1}), \label{eq:maj}
\end{eqnarray}
where $D_{jl}=id_{j,l}d_{j,l+1}$ is defined on each vertical bond.
Since there are no direct connections between any two vertical
bonds, all $D_{jl}$ are\textit{\ good quantum numbers}. Each
$D_{jl}$ acts like a local static $Z_{2}$ gauge field. They commute
with each other and with the Hamiltonian.

Eq. (\ref{eq:maj}) is simply a model of free Majorana fermions with
local Z$ _{2}$ gauge fields. In fact, a similar Majorana fermion
representation of the Hamiltonian was first used by Kitaev
\cite{Kitaev} to solve rigorously the ground state in the
two-dimensional limit. However, in the work of Kitaev,\cite{Kitaev}
a $S=1/2$ spin operator is represented by \textit{four} Majorana
operators, which introduces two redundant degrees of freedom at each
site and many important properties of the system, especially those
involving excitation states, are blurred by those unphysical states.
To remove these redundant degrees of freedom, a complicated
projection on the wave function has to be imposed.

In Eq.(\ref{eq:maj}), $D_{jl}$ for a given $l$ can be all changed
into $-D_{jl}$ by a unitary transformation. For example, the unitary
operator $U_{2l}= \prod_{j}c_{2j,2l}$ can effectively convert all
$D_{2j,2l}$ on the ($2l$)-th ladder to $-D_{2j,2l}$, and
$U_{2l-1}=\prod_{j}c_{2j-1,2l-1}$ converts all $D_{2j-1,2l-1}$ on
the ($2l-1$)-th ladder to $-D_{2j-1,2l-1}$. Therefore, all the
eigenstates of the Hamiltonian are at least $2^{M}$-fold degenerate
for a $M$-leg ladder ($M=0$ for a single chain). In the discussion
below, without loss of generality, we will assume all three coupling
constants $ J_{\alpha }$ ($\alpha =1$, $2$, $3$) to be positive.

\begin{figure}[h]
\centering \includegraphics [width=7cm]{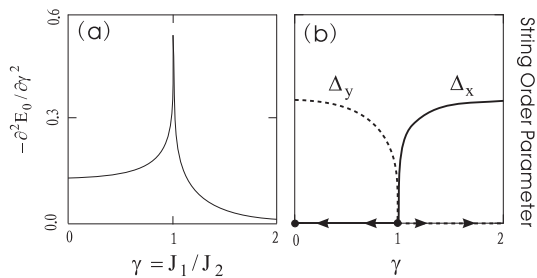}
\caption{The second derivative of the ground state energy density $E_0$ (a)
and the nonlocal string order parameters (b) for a single chain. The arrows
in (b) indicate the RG flow. }
\label{fig:chain}
\end{figure}

For a single spin chain, the Hamiltonian can be readily
diagonalized. The eigen-spectra contain two bands of quasiparticle
excitations. In the ground state, one of the bands is fully
occupied while the other is empty. When $ J_{1}\not=J_{2}$, there
is an energy gap in the low-lying excitations. By varying
$J_{1}/J_{2}$, we find that the second derivative of the ground
state energy density $E_{0}$ diverges logarithmically at
$J_{1}=J_{2}$ where the gap vanishes (Fig. \ref{fig:chain}a).
Furthermore, by applying the renormalization group analysis to
this model, it can be shown that $ J_{1}/J_{2}$ will flow to
infinity if $J_{1}>J_{2}$ or to zero if $ J_{1}<J_{2}$, and
$J_{1}/J_{2}=1$ is a quantum critical point (Fig. \ref
{fig:chain}b).

The transition between the two gapped fermionic states at $J_{1}/J_{2}=1$
does not involve any change of symmetry. Instead it manifests as a change of
topological order. To see this, let us introduce a spin duality
transformation\cite{duality1}
\begin{equation}
\sigma_{j}^{x} = \tau_{j-1}^{x}\tau _{j}^{x},\quad
\sigma_{j}^{y}=\prod_{k=j}^{2N}\tau _{k}^{y}  \label{eq:dual}
\end{equation}
to rewrite the Hamiltonian (\ref{eq:model}) as
\begin{equation}
H_{d}=\sum_{j=1}^{N}\left( J_{1}\tau _{2j-2}^{x}\tau
_{2j}^{x}+J_{2}\tau _{2j}^{y}\right) .  \label{Ising}
\end{equation}
This is nothing but a one-dimensional Ising model with a transverse
field defined in the dual lattice. When $J_{1}>J_{2}$, it is known
that a long range order exists in the dual spin correlation function
of $\tau_{2j}^{x}$ \cite{Pfeuty}
\begin{equation}
\lim_{j\rightarrow \infty }\langle \tau _{0}^{x}\tau
_{2j}^{x}\rangle \sim  \left[ 1-\left( J_{2}/J_{1}\right)
^{2}\right] ^{1/4}. \label{eq:chainx}
\end{equation}
Thus $\tau _{2j}^{x}$ can be regarded as an order parameter
characterizing the phase transition from $J_{1}>J_{2}$ to
$J_{1}<J_{2}$ in the dual space. In the original spin
representation, $\tau _{0}^{x}\tau _{2j}^{x}$ is a string product of
$\sigma _{j}^{x}$:
\begin{equation}
\hat{\Delta}_{x}(j)=\tau _{0}^{x}\tau_{2j}^{x} =
\prod_{k=1}^{2j}\sigma _{k}^{x}=(-1)^{j}\prod_{k=1}^{2j}c_{k}.
\label{eq:strx}
\end{equation}
Eq. (\ref{eq:chainx}) then indicates that $H$ has a hidden
topological order in the $J_{1}>J_{2}$ phase with $\Delta_{x} =
\lim_{j \rightarrow \infty }\langle \hat{\Delta}_{x}(j)\rangle $.

When $J_{1}<J_{2}$, the dual spins $\tau_{2j}^{x}$ become disordered
and $ \Delta _{x}$ vanishes \cite{Pfeuty}. However, by swapping the
$J_{2}$- with the $J_{1}$-term in Eq. (\ref{eq:model}) and applying
a similar duality argument, it can be shown that a string order of
$\sigma _{n}^{y}$
\begin{equation}
\Delta _{y}=\lim_{j\rightarrow \infty }\langle \prod_{k=2}^{2j +
1}\sigma_{k}^{y} \rangle = (-1)^{j}\lim_{j\rightarrow \infty
}\langle \prod_{k=2}^{2j+1}c_{k}\rangle ,  \label{eq:stry}
\end{equation}
is finite in the $J_{1}<J_{2}$ phase and zero otherwise.

The above discussion indicates that in the single-chain limit, the
model is in disordered phases, but it contains two \textit{hidden}
string order parameters. The quantum phase transition at
$J_{1}/J_{2}=1$ corresponds to a continuous change of $\Delta _{x}$
or $\Delta _{y}$ from zero to a finite value from one side of the
critical point to another. In the dual space, however, these
\textit{nonlocal} string order parameters become \textit{local }.
This suggests that Landau-type concepts of continuous phase
transition can still be applied to this simple model, but in the
dual space.

\begin{figure}[tbp]
\centering \includegraphics[width=7cm]{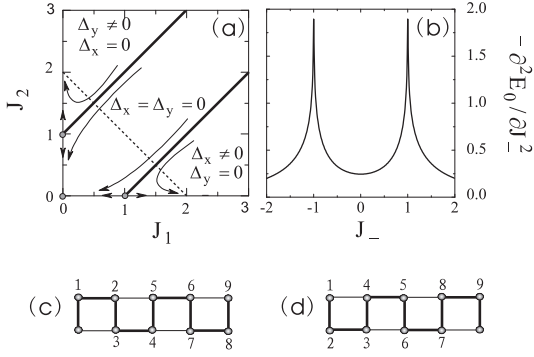} \caption{(a)The
phase diagram with RG flows for the two-leg ladder. (b) $
\partial ^{2}E_{0}/\partial J_{-}^{2}$ ($J_{-}=J_{1}-J_{2}$) along the
dotted line in (a). (c) and (d) are two deformed single chain
representations of the two-leg ladder. $J_{3}$ is set to 1.}
\label{fig:ladder}
\end{figure}

For a two-leg spin ladder, the ground state is in a $\pi $-flux
phase.\cite {flux phase} After diagonalizing the Hamiltonian, we
find that the eigen-spectra contain four branches of fermionic
quasiparticle excitation bands. At zero temperature, two bands are
fully occupied and the other two are completely empty. In general,
there is also an energy gap between the ground and excited states.
But this gap vanishes along both $J_{-}=J_{3}$ and $J_{-}=-J_{3}$
lines ($J_{\pm }=J_{1}\pm J_{2}$). As shown in Fig. (\ref
{fig:ladder}b), the second derivative of the ground state energy
density diverges logarithmically on these two lines.

To clarify the topological nature of these three gapped phases as shown in
Fig (\ref{fig:ladder}a), let us first relabel all the sites along a special
path as shown in Fig. (\ref{fig:ladder}c) to express the original spin
Hamiltonian as an effective single chain model with the third nearest
neighbor couplings:
\begin{equation}
H = \sum_{j}(J_{1}\sigma_{2j-1}^{x}\sigma _{2j}^{x} +
J_{3}\sigma_{2j}^{z} \sigma_{2j+1}^{z} + J_{2}\sigma_{2j}^{y}
\sigma_{2j+3}^{y}). \label{eq:dleg}
\end{equation}
By applying the similar duality transformation (\ref{eq:dual}) to this
Hamiltonian, we find that this model is equivalent to an anisotropic XY spin
chain with a transverse field in the dual space:
\begin{equation}
H_{D}=\sum_{j}(J_{1}\tau_{2j-2}^{x}\tau _{2j}^{x} + J_{2}W_{2j}\tau
_{2j-2}^{y} \tau _{2j}^{y} + J_{3}\tau _{2j}^{z}),
\end{equation}
where $W_{2j}= \tau _{2j-3}^{x}\tau _{2j-1}^{z} \tau _{2j+1}^{x}$ is
the plaquette operator defined in the dual space. It is a good
quantum number.

In the ground state, $W_{2j}=-1$, corresponding to the $\pi $-flux
phase. It is known that the system is in a long-range ordered state
of $\tau _{2j}^{x}$ when $J_{-}>J_{3}$ and in a disordered state
otherwise.\cite{McCoy} Thus, $\langle \tau _{2j}^{x}\rangle$ is an
effective order parameter characterizing the quantum phase
transition across the critical line $ J_{-}=J_{3}$. Back to the
original spin model, there exists a string order parameter $\Delta
_{x}$ defined by Eq. (\ref{eq:strx}) but in the deformed lattice
(Fig. \ref{fig:ladder}c) in the $J_{-}>J_{3}$ phase. Using the
results given in Ref. \cite{McCoy}, we find that
\begin{equation}
\Delta _{x}\sim \frac{\sqrt{J_{+}/J_{-}}}{2(1+J_{+}/J_{-})}\left[ 1-\left(
\frac{J_{3}}{J_{-}}\right) ^{2}\right] ^{1/4}.
\end{equation}
In the other two phases ($J_{-}<J_{3}$), $\Delta _{x}=0$.

The above duality argument can be generalized to clarify the quantum
phase transition across the line $J_{-}=-J_{3}$ by swapping $J_{1}$-
with $J_{2}$ -term in the original Hamiltonian. In the ground state,
it can be also shown that the nonlocal string order parameter
$\Delta _{y}$ defined by Eq.(\ref {eq:stry}), but now in a deformed
lattice where $\sigma _{j}^{y}$ is labelled according to Fig.
(\ref{fig:ladder}d), is finite in the phase $ J_{-}<-J_{3}$, and
vanishes in the other two phases. Thus the phase diagram of the
two-leg ladder system can be classified by the two nonlocal string
order parameters: $\Delta _{x}$ and $\Delta _{y}$. They are finite
in the $ J_{-}>J_{3}$ and $J_{-}<-J_{3}$ phases, respectively. The
quantum phase transition at each critical line $J_{-}=\pm J_{3}$
corresponds to a continuous change of one of the string order
parameters from zero to a finite value.

For a $2M$-row brick wall lattice with $M>1$, the ground state of
the model is in a zero-flux phase \cite{flux phase} and all $D_{jl}$
for a given $l$ have the same value. The ground state is thus
$2^{2M}$-fold degenerate, and this massive degeneracy leads to an
extensive zero temperature entropy. In what follows, we will focus
on the case $D_{jl}=(-)^{l}$. The resulting conclusions can be
easily generalized to other degenerate states by unitary
transformations.

\begin{figure}[tbp]
\centering\includegraphics [width=6cm] {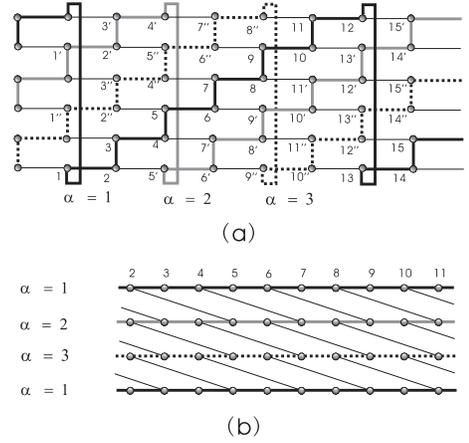}
\caption{(a) The six-row brick wall lattice and (b) its equivalent three-row
lattice. }
\label{fig:leg6}
\end{figure}

To elucidate the topological nature of the quantum criticality in this
system, let us first follow what we have done for the two-leg ladder to map
a $2M$-row lattice to a $M$-row lattice by chaining all neighboring sites
along the stair paths of $J_{1}-J_{3}$ links with periodic boundary
conditions. An example of such mappings is displayed in Fig. \ref{fig:leg6}
for a six-row brick wall lattice. In the new lattice system, the Hamiltonian
(\ref{eq:maj}) can be expressed as
\begin{eqnarray}
H &=& -i\sum_{j=1}^{2N}\sum_{\alpha =1}^{M}(J_{1}c_{2j-1,\alpha }
c_{2j,\alpha } - J_{2}c_{2j,\alpha } c_{2j+3,\alpha +1}  \notag \\
&&\text{ \ \ \ \ \ \ \ \ \ \ } + J_{3}(-)^{j}c_{2j,\alpha }
c_{2j+1,\alpha }). \label{eq:ham1}
\end{eqnarray}
By transforming the second index of $c_{j,\alpha }$ into its
momentum form $c_{j, q}$, Eq.(\ref{eq:ham1}) then becomes $H =
\sum_{q} H_{q}$, where
\begin{eqnarray}
H_{q} &=&-i\sum_{j}
[J_{1} c_{2j-1,-q} c_{2j,q}-J_{2}e^{iq}c_{2j,-q}c_{2j+3, q} \notag \\
&&\text{ \ \ \ \ \ } + J_{3}(-)^{j} c_{2j,-q} c_{2j+1,q}],
\label{eq:legM}
\end{eqnarray} and
$q=2\pi m/M$ and $m=0\cdots M-1$. This Hamiltonian is now block
diagonalized according to the value of $|q|$. When $q=0$ or $\pi $,
$-q$ is equal or equivalent to $q$, and $c_{j,q}$ is still a
Majorana fermion. In other cases, $c_{j,-q}$ and $c_{j,q}$ are
conjugate pairs of standard fermion operators, i.e. $c_{j,-q} =
c_{j,q}^{\dagger }$.

\begin{figure}[tbp]
\includegraphics [width=8cm] {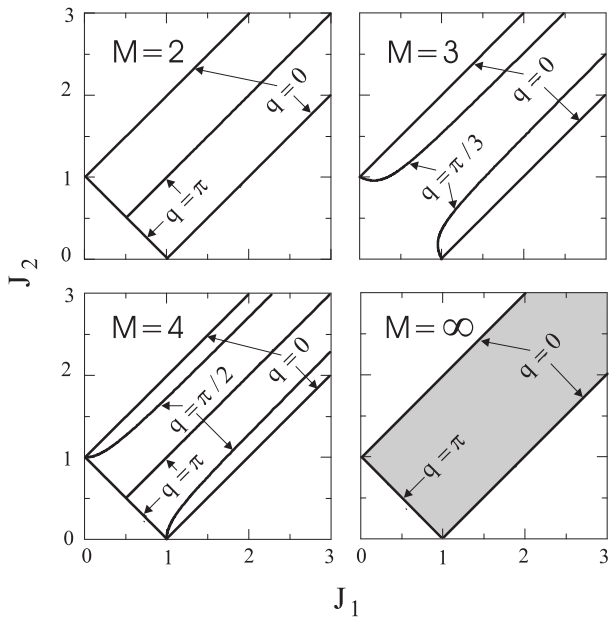}
\caption{The phase diagram with quantum critical lines of the
$2M$-row brick-wall lattice. $J_3= 1$ } \label{fig:phasedia}
\end{figure}

The quasiparticle eigen-spectra can be obtained by diagonalizing the
Hamiltonian at each $|q|$ sector ($|q|\leq \pi $ ). Fig.
\ref{fig:phasedia} shows the phase diagram of the multi-row lattices
with $M=2$, $3$, $4$ and $ \infty $. For a $2M$-row lattice, there
are $M+2$ ($M+1$) critical lines if $ M=\text{ even}$
($M=\text{odd}$). These critical lines correspond to the vanishing
gap lines of quasiparticle excitations, where the second derivatives
of the ground state energy density still diverge logarithmically.

When $q=0$, $H_{q}$ has exactly the same form as the Hamiltonian of
a two-leg ladder with a $\pi $-flux, represented in the deformed
lattice in Fig. (\ref{fig:ladder}c). As discussed before, the
critical excitation gives rise to two quantum phase transitions at
the lines $J_{-} = \pm J_{3}$. The corresponding topological order
parameters, $\Delta _{x,0}$ and $\Delta _{y,0}$, can be defined from
Eqs.(\ref{eq:strx},\ref{eq:stry}) by replacing $c_{j}$ with
$c_{j,q=0}$. $\Delta_{x,0}$ is finite in the phase $J_{-}>J_{3}$ ,
while $\Delta _{y,0}$ is finite in the phase $J_{-}<-J_{3}$.

When $q=\pi $ ($M=\text{even}$ only), $H_{q}$ is also equivalent to
the Hamiltonian of a two-leg ladder, but with a minus $J_{2}$-term.
The ground state of $H_{q}$ is effectively in zero-flux. Two
critical lines are now located at $J_{+}=J_{3}$ and at $J_{-}=0$
with $J_{+}>J_{3}$. Correspondingly, one can also define two
topological string order parameters, $\Delta _{x,\pi }$ and
$\Delta_{y,\pi }$, from Eqs. (\ref {eq:strx},\ref{eq:stry}) by
replacing $c_{j}$ with $c_{j,q=\pi }$. $\Delta _{x,\pi }$ ($\Delta
_{y,\pi }$) is finite in the $J_{+}>J_{3}$ and $ J_{1}>J_{2}$
($J_{1}<J_{2}$) phase.

When $q\not=0$ or $\pi $, $c_{j, \pm q}$ are no longer Majorana
fermions. By considering the limiting cases of $J_{1}\gg J_{2}$ and
$J_{1}\ll J_{2}$, it can be shown that two string order parameters
defined from the combinations of $c_{j, q}$ and $c_{j, -q}$ exists.
However, as $H_{q}$ is coupled with $ H_{-q}$, we are still unable
to write down accurately the expressions of the string order
parameters.

In the two dimensional limit ($M=\infty $), all the critical lines
inside the regime enclosed by the three curves, $J_{-}=\pm J_{3}$
and $J_{+}=J_{3}$ , merge together and form a continuum gapless
phase. Therefore, the whole phase space contains three gapped phases
and one gapless phase, in agreement with the phase diagram obtained
by Kitaev.\cite{Kitaev} The gapless phase may have a complicated
non-Abelian topological structure.\cite{Kitaev} However, the phase
transition from any gapped phase to the gapless phase can be
characterized by a string order parameter, since the phase
boundaries of the gapless phase are fully determined by the phase
transition lines of the decoupled fermionic chains with $q=0$ and
$\pi $.

In summary, using the Jordan-Wigner transformation, we have shown
that the Hamiltonian defined by Eq.(\ref{eq:model}) is mapped onto a
model of free Majorana fermions with local $Z_{2}$ gauge fields
\textit{without} any redundant degrees of freedom. By solving this
model \textit{exactly} at zero temperature, we find that the system
undergoes a number of continuous phase transitions. These quantum
phase transitions are not induced by spontaneous symmetry breaking
and there are no conventional Landau-type local order parameters.
However, each transition corresponds to a continuous change of a
topological string order parameter from a finite value to zero. This
reveals that the quantum phase transitions in this system can not
only be indexed by a topological quantum number, but also be
characterized by a topological order parameter. In the dual space,
the string order parameter actually corresponds to a \textit{local}
order parameter and the basic concepts of Landau theory of
continuous phase transition are still applicable.

The above conclusion is drawn based on the analysis of a specific
model (\ref{eq:model}), however, we believe it might be valid
generically. Similar topological string order parameters have been
found, for example, in the Haldane gapped spin
chains.\cite{nijs-rommelse,kennedy-tasaki} Further exploration of
this problem is desired, which may lead to a unified description of
convention and topological quantum phase transitions.

We thank L. Yu for helpful discussions. Support from the NSFC and
the national program for basic research is acknowledged.

\end{document}